\documentclass{article}

\usepackage[nonatbib,preprint]{neurips_2022_modified}

\usepackage{amsmath}
\usepackage{graphicx}
\usepackage{array}
\usepackage{color}
\usepackage{amsfonts}
\usepackage{booktabs} 
\usepackage{hyperref}
\usepackage{multirow}
\usepackage{caption}
\usepackage{booktabs} 
\usepackage[normalem]{ulem}
\useunder{\uline}{\ul}{}
\usepackage{tikz}
\usetikzlibrary{trees,shapes} 
\usepackage{listings} 
\usepackage{mdframed} 
\usepackage{fancyvrb} 
\usepackage{wrapfig}
\usepackage{subcaption}
\usepackage{framed}
\usepackage{adjustbox}
\usepackage[T1]{fontenc}
\usepackage{colortbl}
\usepackage[misc,geometry]{ifsym}
\usepackage{cite} 
\usepackage{xspace} 

\definecolor{skyblue}{rgb}{0.53, 0.81, 0.92}

\definecolor{keywordcolor}{rgb}{0.0, 0.0, 1.0}
\definecolor{tcolor}{rgb}{0.5, 0.0, 0.0}

\lstdefinestyle{dslstyle}{
    basicstyle=\ttfamily,
    keywordstyle=\color{keywordcolor}\bfseries,
    identifierstyle=\color{tcolor},
    columns=fullflexible,
}

\newcommand{\textcode}[1]{{\fontfamily{cmtt}\selectfont #1}\xspace} 

\newcommand{\task}{\texttt{{T}}}
\newcommand{\code}{\texttt{{C}}}
\newcommand{\goal}{\texttt{G}}
\newcommand{\cons}{{\texttt{L}}}
\newcommand{\grid}{\texttt{W}}
\newcommand{\sketch}{\texttt{S}}

\newcommand{\fd}{\texttt{forward}}
\newcommand{\bk}{\texttt{back}}
\newcommand{\lt}{\texttt{left}}
\newcommand{\rt}{\texttt{right}}
\newcommand{\setpc}{\texttt{setpencolor}}
\newcommand{\Repeat}{\texttt{repeat}}
\newcommand{\atmost}{\texttt{AtMost}}
\newcommand{\startby}{\texttt{StartBy}}
\newcommand{\exactly}{\texttt{Exactly}}
\newcommand{\none}{\texttt{None}}

\newcommand{\inp}{\texttt{in}}
\newcommand{\out}{\texttt{out}}

\newcommand{\find}{\texttt{Find}}
\newcommand{\findonly}{\texttt{FindOnly}}
\newcommand{\findforb}{\texttt{FindForbid}}

\newcommand{\counting}{\texttt{Count}}

\newcommand{\draw}{\texttt{Draw}}

\newcommand{\medium}{\texttt{medium}}

\newcommand{\Easy}{\texttt{Easy}}
\newcommand{\Medium}{\texttt{Medium}}
\newcommand{\Hard}{\texttt{Hard}}

\newcommand{\diff}{\texttt{D}}

\newcommand{\expertsyn}{\textsc{ExpertSyn}}
\newcommand{\rotateflip}{\textsc{RotateFlip}}
\newcommand{\humansyn}{\textsc{HumanSyn}}
\newcommand{\ours}{\textsc{XLogoSyn}}

\newcommand{\mini}{\text{XLOMini}}

\title{Task Synthesis for Elementary Visual Programming in XLogoOnline Environment
}

\author{
    \textbf{Chao Wen}\textsuperscript{1} \quad
    \textbf{Ahana Ghosh}\textsuperscript{1} \quad
    \textbf{Jacqueline Staub}\textsuperscript{2}\quad
    \textbf{Adish Singla}\textsuperscript{1} \\
    \textsuperscript{1}MPI-SWS, \textcode{\{chaowen, gahana, adishs\}@mpi-sws.org}\\ 
    \textsuperscript{2}University of Trier, \textcode{staub@uni-trier.de}
}

\begin{document}
\maketitle   



\begin{abstract} 
\looseness-1In recent years, the XLogoOnline programming platform has gained popularity among novice learners. It integrates the Logo programming language with visual programming, providing a visual interface for learning computing concepts.
However, XLogoOnline offers only a limited set of tasks, which are inadequate for learners to master the computing concepts that require sufficient practice. 
To address this, we introduce \ours{}, a novel technique for synthesizing high-quality tasks for varying difficulty levels. Given a reference task, \ours{} can generate practice tasks at varying difficulty levels that cater to the varied needs and abilities of different learners.
\ours{} achieves this by combining symbolic execution and constraint satisfaction techniques. Our expert study demonstrates the effectiveness of \ours{}. We have also deployed synthesized practice tasks into XLogoOnline, highlighting the educational benefits of these synthesized practice tasks.
\end{abstract}



\section{Introduction}\label{sec.intro}

In recent years, XLogoOnline~\cite{DBLP:conf/issep/HromkovicSS17,xlogoonline} has emerged as a new platform, which uniquely integrates the traditional Logo programming language~\cite{pea1987logo} with the visual programming paradigm. XLogoOnline has been adopted in hundreds of educational courses and is utilized by tens of thousands of students every year~\cite{DBLP:conf/issep/HromkovicSS17,DBLP:journals/eatcs/Staub21}.
XLogoOnline~\cite{DBLP:conf/issep/HromkovicSS17} is organized into four programming levels: \textit{Mini}, \textit{Midi}, \textit{Maxi}, and \textit{Mega}, each offering tasks tailored to specific age groups. 

We focus on the \textit{Mini} level (referred to as \mini{}), which centers around problem-solving, incorporating computing concepts like loops and basic mathematics. In \mini{}, learners are given tasks and a few code blocks, including basic commands \fd{}, \bk{}, \lt{}, \rt{}, the state-based command \setpc{}, and the control structure \Repeat{}.
Each task contains a visual grid with a turtle, descriptive text outlining the goal of the task, and code constraints. Learners must construct the code satisfying the code constraints and then execute it to direct the turtle's movement to achieve the goal. 
%
\begin{figure}[t!]
    \centering
    \includegraphics[width=0.95\linewidth]{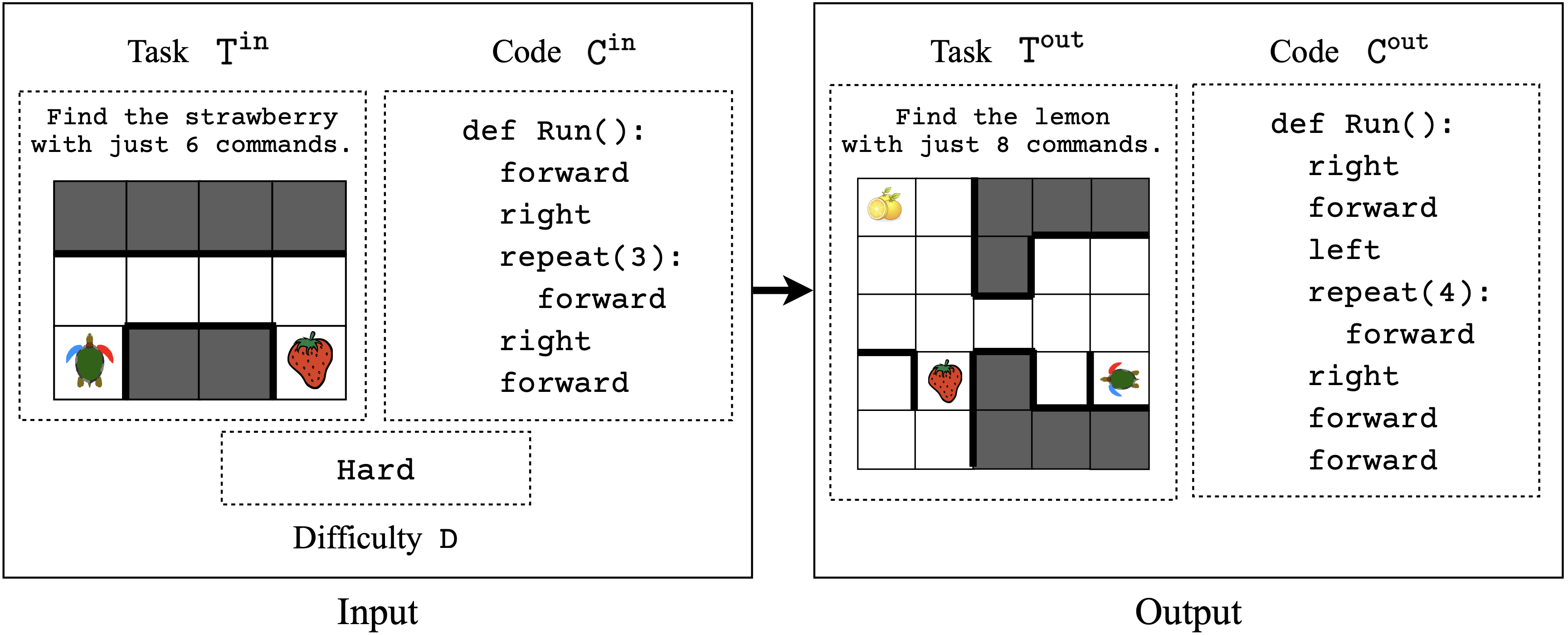}
    
    \caption{Illustration of \ours{} for reference task $87$ from \mini{}~\cite{xlogoonline}. \ours{}'s input includes a reference task $\task^\inp$, its solution code $\code^\inp$, and the desired difficulty level $\diff$ for a new practice task. The output includes a generated task $\task^\out$ and its solution code $\code^\out$ that satisfies the desired difficulty level w.r.t. the reference task.}
    \label{fig:input-output-example}
\end{figure}
%
However, \mini{} only offers a limited set of tasks. The scarcity of tasks may hinder learners from mastering computing concepts such as loops and basic mathematics, which require sufficient practice to deepen learners' understanding.

To address this, we propose \ours{}, a technique for synthesizing high-quality practice tasks at a specific difficulty level. Fig.~\ref{fig:input-output-example} shows an example of the input and output of \ours{}. Given a reference task, its solution code, and the desired difficulty, our technique can generate numerous varied tasks, addressing task scarcity and eliminating the need for manually crafting tasks. 

\looseness-1Our key contributions include: (i) We develop \ours{}, a technique to automatically generate high-quality tasks at a desired difficulty level; (ii) We conduct an expert study to show that \ours{} can synthesize tasks with a quality close to those crafted by experts; (iii) We deploy synthesized practice tasks by \ours{} on XLogoOnline and report on initial results highlighting the educational benefits.\footnote{Implementation of \ours{} is publicly available at: \\ \text{\quad \ \ \  }\url{https://github.com/machine-teaching-group/aied2024-xlogo-tasksyn}}\label{footnote:link}

%



\subsection{Related Work}\label{sec.related-work}

{\textbf{Content generation.}} Content generation has been explored in domains such as game content generation~\cite{DBLP:journals/tciaig/SmithM11,DBLP:conf/aied/ParkMMWBL20} and math problem generation~\cite{DBLP:conf/ijcai/PolozovOSZGP15,DBLP:conf/aaai/AlvinGMM14}. Existing works often create a template filled with placeholders, followed by a search within this template for solutions~\cite{DBLP:journals/tciaig/SmithM11,DBLP:conf/aied/ParkMMWBL20,DBLP:conf/ijcai/PolozovOSZGP15}. To search for solutions, most works first encode the problem using a logic representation and then use Answer Set Programming solvers~\cite{DBLP:journals/tciaig/SmithM11,DBLP:conf/aied/ParkMMWBL20} or Satisfiability Modulo Theories (SMT) solvers~\cite{DBLP:conf/tacas/MouraB08}. We employ SMT solvers; however, we further combine SMT solvers with symbolic code execution to navigate the search for solutions within the visual task grid.

\looseness-1{\textbf{Task synthesis in visual programming.}} Recent works have studied task synthesis in visual programming domains. Most approaches utilize constraint solving techniques and symbolic code execution~\cite{DBLP:conf/nips/AhmedCEFGRS20,DBLP:conf/aied/GhoshTDS22}, with some recent works also incorporating reinforcement learning-based strategies to speed up the generation process~\cite{DBLP:journals/corr/abs-2305-18342}. However, these works have not considered the \mini{} domain, which is characterized by a diverse range of task types, grid elements, and state-based commands (e.g., \setpc{} in \mini{}). Furthermore, existing techniques have not incorporated task difficulty into the generation process.

{\textbf{Large language models for programming task synthesis.}} Recent works have explored large language models (LLMs) to synthesize tasks in programming domains~\cite{chatgpt,DBLP:journals/corr/abs-2402-01580}. 
Existing works have shown LLMs' potential in generating and solving tasks for text-based programming domains such as Python~\cite{DBLP:conf/icer/SarsaDH022,DBLP:journals/corr/abs-2306-17156}. 
However, state-of-the-art LLMs still struggle in visual programming as they are unable to combine spatial, logical, and programming skills~\cite{DBLP:journals/corr/abs-2305-18342,singla2023evaluating}.


\section{Preliminaries and Problem Setup}\label{sec.setup}

\begin{figure}[t!]
    \centering
    \setlength{\FrameSep}{3pt}
    \captionsetup[subfigure]{skip=10pt}
    \captionsetup[subfigure]{belowskip=-6pt}
    \begin{subfigure}{0.32\textwidth}
        \caption*{\textcolor{blue}{38 (\counting)}: Collect exactly 10 strawberries.}
        \centering
        \begin{mdframed}[innerleftmargin=2pt,innerrightmargin=2pt,innertopmargin=-3pt,innerbottommargin=-3pt]
        \begin{minipage}{0.48\linewidth}
            \includegraphics[width=\linewidth]{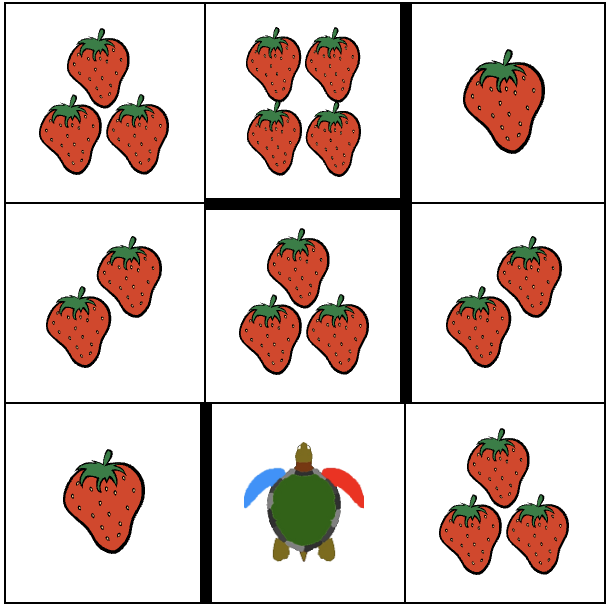}
        \end{minipage}
        \hfill
        \begin{minipage}{0.45\linewidth}
            \begin{lstlisting}
def Run():
  forward
  left
  forward
  back
  left
  forward
  ...
            \end{lstlisting}
        \end{minipage}
    \end{mdframed}
    \end{subfigure}
    \hfill
    \vspace{-5pt}
    \begin{subfigure}{0.32\textwidth}
        \centering
        \caption*{\textcolor{blue}{54 (\draw)}: Draw the picture without ``forward''.}
        \begin{mdframed}[innerleftmargin=2pt,innerrightmargin=2pt,innertopmargin=-3pt,innerbottommargin=-3pt]
        \begin{minipage}{0.40\linewidth}
            \includegraphics[width=\linewidth]{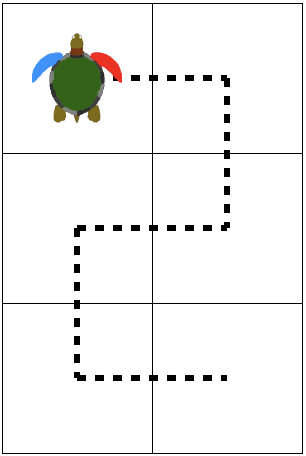}
        \end{minipage}
        \hfill
        \begin{minipage}{0.45\linewidth}
            \begin{lstlisting}
def Run():
  left
  back
  repeat(2):
    right
    back
  repeat(2):
    ...
            \end{lstlisting}
        \end{minipage}
        \end{mdframed}
    \end{subfigure}
    \hfill
    \begin{subfigure}{0.32\textwidth}
        \centering
        \caption*{\textcolor{blue}{87 (\find)}: Find the strawberry with just 6 commands.}
    \begin{mdframed}[innerleftmargin=2pt,innerrightmargin=2pt,innertopmargin=-3pt,innerbottommargin=-3pt]
        \begin{minipage}{0.53\linewidth}
            \includegraphics[width=0.95\linewidth]{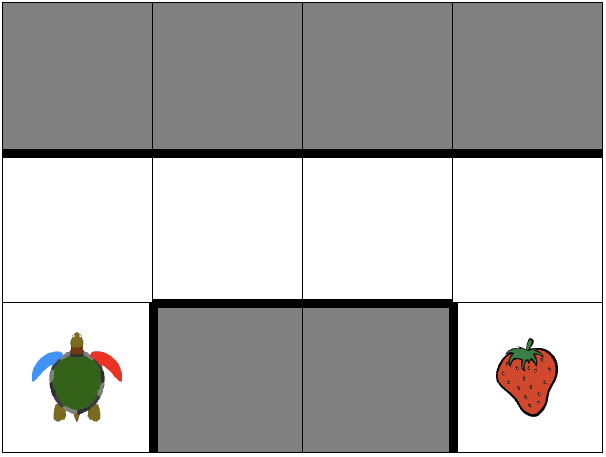}
        \end{minipage}
        \hfill
        \begin{minipage}{0.45\linewidth}
            \begin{lstlisting}
def Run():
  forward
  right
  repeat(3):
    forward
  right
  forward

            \end{lstlisting}
        \end{minipage}
    \end{mdframed}
    \end{subfigure}
    \caption{Illustrative examples of reference tasks and their solution codes in \mini{}.}
    \label{fig:goal_types}
\end{figure}

{\textbf{Task and code specifications in \mini{}.}} A task $\task := (\goal, \cons, \grid)$ includes a goal $\goal$ (the turtle's objective), code constraints $\cons$ (constraints for solution code), and a visual grid world $\grid$ (a two-dimensional grid with a turtle and elements like fruits and walls). To solve a task, a learner writes a \emph{solution code} that meets the task's code constraints and achieves the goal when executed on the visual grid world. Fig. \ref{fig:goal_types} shows examples of tasks and their solution codes in \mini{}.



{\textbf{Levels of task difficulty.}} Given a reference task $\task^\inp$, we define the relative difficulty $\diff$ of a new practice task $\task^\out$ as follows: (i) \Easy{}: Solving $\task^\out$ requires no additional concepts or steps beyond what is required for solving $\task^\inp$; (ii) \Medium{}: Solving $\task^\out$ requires additional steps and understanding of concepts beyond those required to solve $\task^\inp$; (iii) \Hard{}: Solving $\task^\out$ requires additional steps and understanding of concepts beyond those required to solve a \medium{} task of $\task^\inp$. 
In Section~\ref{sec:method}, we provide concrete criteria for these difficulty levels within the \mini{} domain, as part of the implementation details for \ours{}.

{\textbf{Evaluation of synthesized tasks.}} We use a multidimensional rubric to assess the quality of a \textit{synthesized} task-code pair $(\task^\out, \code^\out)$ for a \textit{reference} task-code pair $(\task^\inp, \code^\inp)$ at a specified difficulty level $\diff$. 
The rubric consists of the following five metrics: 
(i) \emph{Visual quality} evaluates the distinctiveness and aesthetic appeal of $\task^\out$; 
(ii) \emph{Concept similarity} evaluates the alignment of concepts between the $(\task^\out, \code^\out)$ and $(\task^\inp, \code^\inp)$; 
(iii) \emph{Elements utility} evaluates the usefulness of grid elements in $\task^\out$; 
(iv) \emph{Code quality} evaluates the correctness of $\code^\out$; 
(v) \emph{Difficulty consistency} evaluates the consistency of the difficulty of $\task^\out$ w.r.t. the difficulty level $\diff$. 
Domain experts rate each metric on a three-point Likert scale: $0$ for low quality, $0.5$ for acceptable quality, and $1$ for excellent quality. 
The \emph{overall quality} of $\task^\out$ is defined as the minimum rating across the five metrics.

\looseness-1{\textbf{Task synthesis objective.}} Given a reference task $\task^\inp$, its corresponding solution code $\code^\inp$, and a specified difficulty level $\diff$, our objective is to automatically synthesize a set of high-quality task-code pairs $\{(\task^{\out}, \code^{\out})\}$ (see Fig.~\ref{fig:input-output-example}).


\section{Our Task Synthesis Technique: \ours{}}\label{sec:method}

\begin{figure}[t!]
    \centering
    \begin{subfigure}{\textwidth}
        \centering
        \includegraphics[width=1\linewidth]{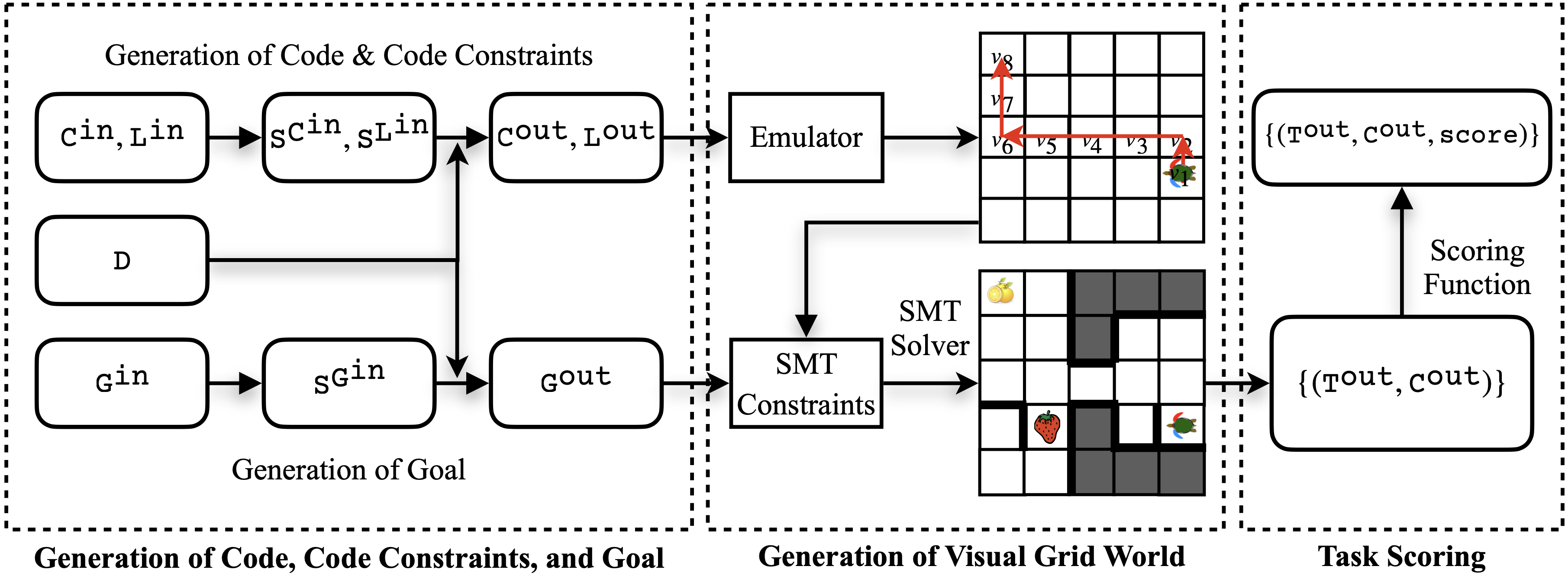}
        \caption{Stages of our task synthesis technique \ours{}}
        \label{fig:technique_stages}
    \end{subfigure}
    \begin{subfigure}{1\textwidth}
        \centering
        \begin{minipage}[tb]{0.53\linewidth}
          \centering
          \begin{minipage}[tb]{0.32\linewidth}
          \begin{mdframed}[innerleftmargin=3pt,innerrightmargin=3pt,innertopmargin=-4pt,innerbottommargin=-4pt]
          \begin{lstlisting}
def Run():
  forward
  right
  repeat(3):
    forward
  right
  forward



          \end{lstlisting}
          \end{mdframed}
          \vspace{-12pt}
          \end{minipage}%
          \hfill
          \begin{minipage}[tb]{0.32\linewidth}
            \centering
        \begin{mdframed}[innerleftmargin=3pt,innerrightmargin=3pt,innertopmargin=-4pt,innerbottommargin=-4pt]
        \begin{lstlisting}[]
def Run():
  [B1]
  [B2]
  [B3]
  repeat([X]):
    [B4]
    [B5]
  [B6]
  [B7]
  [B8]
        \end{lstlisting}
        \end{mdframed}
        \vspace{-12pt}
          \end{minipage}%
          \hfill
          \begin{minipage}[tb]{0.32\linewidth}
            \centering
        \begin{mdframed}[innerleftmargin=3pt,innerrightmargin=3pt,innertopmargin=-4pt,innerbottommargin=-4pt]
            \begin{lstlisting}
def Run():
  right
  forward
  left
  repeat(4):
    forward
  right
  forward
  forward

            \end{lstlisting}
            \end{mdframed}
            \vspace{-12pt}
          \end{minipage}%
          \caption{$\code^{\inp}, \sketch^{\code^{\inp}}, \code^{\out}$}
          \label{fig:technique.cin-scin-cout}
        \end{minipage}
        \hfill
        \begin{minipage}[tb]{0.45\linewidth}
            \centering
            \vspace{-2pt}
            \begin{minipage}[tb]{1\linewidth}
              \centering
              \begin{mdframed}[innerleftmargin=2pt,innerrightmargin=-1pt,innertopmargin=-4pt,innerbottommargin=-4pt]
              \begin{lstlisting}[mathescape=true]
$\cons^\inp\;\;$ = "Use just 6 commands"
$\sketch^{\cons^{\inp}}$ = "Use [cons_type] [N] commands"
$\cons^{\out}\:$ = "Use just 8 commands"
              \end{lstlisting}
            \end{mdframed}
            \vspace{-14pt}
            \caption{$\cons^{\inp}, \sketch^{\cons^{\inp}}, \cons^{\out}$}
            \label{fig:technique.cons}
            \end{minipage}
            \vfill
            \begin{minipage}[tb]{1\linewidth}
              \centering
              \vspace{-2pt}
              \begin{mdframed}[innerleftmargin=2pt,innerrightmargin=-1pt,innertopmargin=-4pt,innerbottommargin=-4pt]
              \begin{lstlisting}[mathescape=true]
$\goal^\inp\;\;$ = "Find the strawberry"
$\sketch^{\goal^{\inp}}$ = "[task_type] the [fruit_type]"
$\goal^{\out}\;$ = "Find the lemon"
              \end{lstlisting}
            \end{mdframed}
              \vspace{-14pt}
              \caption{$\goal^{\inp}, \sketch^{\goal^{\inp}}, \goal^{\out}$}
              \label{fig:technique.goal}
            \end{minipage}
        \end{minipage}
    \end{subfigure}
    \caption{
    (a) illustrates the stages of \ours{}. 
    (b)--(d) show examples of different components after applying these stages to Fig.~\ref{fig:input-output-example} (Input). 
    Specifically, 
    (b) shows the input code, its sketch, and the output code, where \texttt{B1}, \texttt{B2}, $\cdots$, \texttt{B8} $\in$ \{\none{}, \lt{}, \rt{}, \ldots\} and \texttt{X} $\in$ \{2, 3, \ldots\}. 
    (c) shows the input code constraints, its sketch, and the output code constraints, where \(\texttt{cons\_type}\) $\in$ \{\atmost{}, \exactly{}, \startby{}, \none{}\} and \texttt{N} $\in$ \{1, 2,\ldots\}. 
    (d) shows the input goal, its sketch, and the output goal, where \(\texttt{task\_type}\) $\in$ \{\find, \findonly, \findforb, ...\} and \texttt{fruit\_type} $\in$ \{\texttt{strawberry}, \texttt{lemon}\}. 
}
    \label{fig:technique}
    \vspace{-1mm}
\end{figure}

In this section, we provide an overview of our task synthesis technique \ours{}. Fig.~\ref{fig:technique_stages} illustrates the three stages of \ours{}. 

\setlength{\FrameSep}{-2pt} 

{\textbf{Stage 1: Generation of code, code constraints, and goal.}}\label{sec:method.code_goal_generation} 
Fig.~\ref{fig:technique_stages} illustrates the first stage of \ours{}. This stage first creates a template for each of $\code^\inp$, $\cons^\inp$, and $\goal^\inp$~\cite{DBLP:conf/nips/AhmedCEFGRS20,DBLP:conf/aied/GhoshTDS22}. The templates express the high-level structures while leaving low-level details unspecified with placeholders. 
Then, we fill in the placeholders with specific values using an SMT-based constraint solver~\cite{DBLP:conf/tacas/MouraB08}. For example, in Figs.~\ref{fig:technique.cin-scin-cout}, ~\ref{fig:technique.cons}, and~\ref{fig:technique.goal}, placeholder \texttt{[B1]} is replaced with ``\texttt{right}'' and \texttt{[fruit\_type]} with ``\texttt{lemon}''. During instantiation, we also incorporate SMT constraints based on the input difficulty $\diff$. The difficulty $\diff$ controls the difficulty of the generated outputs. At $\diff = \Easy{}$, we maintain the original code length without extra constraints. At $\diff = \Medium$, we allow code sequences up to $2$ commands longer than $\code^\inp$. At $\diff = \Hard{}$, the code sequences need to be exactly $2$ commands longer than $\code^\inp$ and we allow an extra code constraint. 
Moreover, $\goal^{\out}$ is allowed to differ from $\goal^{\inp}$ only at $\diff = \Hard{}$. After this stage, we obtain $(\code^{\out}, \cons^{\out}, \goal^{\out})$, aligned with the specified difficulty level $\diff$.

\looseness-1{\textbf{Stage 2: Generation of visual grid world.}}\label{subsec.task_grid_gen} This stage synthesizes a visual grid world $\grid^\out$ using the outputs $(\code^\out, \cons^\out, \goal^\out)$ from the previous stage. First, we create an empty grid and randomly initialize the turtle's starting location and direction (see Fig.~\ref{fig:technique_stages}). Then, we symbolically execute code $\code^{\out}$ on the empty grid using an emulator, producing a trajectory of visited grid cells $(v_1, v_2, \cdots, v_n)$, highlighted in red in Fig.~\ref{fig:technique_stages}. Next, we use the goal $\goal^{\out}$ and the trajectory to formulate SMT constraints concerning the placement of various grid elements such as fruits, walls, etc. For example, if the goal is ``Find the strawberry'', a strawberry must be placed in the final grid cell $v_n$. An SMT solver solves these constraints to generate a visual grid world $\grid^{\out}$. Finally, we merge $\grid^\out$ with the outputs from the previous stage ($\code^\out$, $\cons^\out$, $\goal^\out$), to obtain output task-code pairs $(\task^{\out}, \code^{\out})$, where $\task^\out = ({\goal^{\out}}, {\cons^{\out}}, {\grid^{\out}})$.

\looseness-1{\textbf{Stage 3: Task scoring.}} \label{subsec.task_scoring} In the final stage, we apply a scoring function to evaluate the quality of task-code pairs $\{(\task^{\out}, \code^{\out})\}$, inspired by scoring functions considered in literature on task synthesis for visual programming~\cite{DBLP:conf/nips/AhmedCEFGRS20,DBLP:conf/aied/GhoshTDS22}. 


\section{Evaluation of \ours{} Using Expert Study}\label{sec.exp2}

{\textbf{Techniques evaluated.}} We compare \ours{} with three different techniques. Each technique accepts an input specification ($\task^{\inp}, \code^{\inp}, \diff$), which includes the reference task, its corresponding solution code, and the desired difficulty level respectively, and generates the output task. We consider the following baselines:

\begin{enumerate}
    \item \expertsyn{} involves an expert in \mini{} carefully crafting a task \(\task^{\out}\) and its code \(\code^{\out}\) based on the input specification.
    \item \humansyn{} uses a collection of 1,331 user-created tasks from XLogoOnline to create tasks and solution codes~\cite{DBLP:conf/issep/HromkovicSS17}. During the creation of this collection, users created their tasks without specific reference tasks to guide their synthesis process. Given this collection and the input specification, this technique generates a task and its solution code as follows: the expert (the same as the one in \expertsyn{}) selects a task $\task^\out$ that matches the input specification from the collection, considering only tasks of the same type as $\task^{\inp}$. After selecting $\task^\out$, the expert crafts an optimal solution code \(\code^{\out}\) for \(\task^{\out}\).
    \item \rotateflip{} generates tasks by applying rotations and flips to the input task's grid. For \diff{} = \Easy{}, it rotates the grid 90 degrees counterclockwise without altering \(\code^{\out}\). For \diff{} = \Medium{}, it performs a mirror flip of the grid and adjusts \(\code^{\out}\) to match the flipped grid. For \diff{} = \Hard{}, it performs both rotation and mirror flip for \(\task^\out\) and adjusts \(\code^{\out}\) as needed.
\end{enumerate}

{\textbf{Experimental setup.}} We selected 24 reference tasks from \mini{} covering a broad range of concepts. For these reference tasks, we generated practice tasks at \Easy{}, \Medium{}, and \Hard{} levels with each technique. 
We considered $288$ scenarios ($24$ reference tasks $\times$ $3$ difficulty levels $\times$ $4$ techniques). However, for \humansyn{}, $15$ scenarios of \draw{} type were missing from the collection of user-created tasks ($5$ \draw{} tasks $\times$ $3$ difficulty levels), resulting in a final count of $273$ scenarios. Two independent human evaluators, not involved in \expertsyn{} or \humansyn{}, scored each scenario based on the rubric in Section~\ref{sec.setup}.
The final score for a scenario was derived by averaging the scores provided by the two evaluators aggregated over different dimensions. During evaluation, the origin of each scenario was hidden and scenarios were presented in a randomized order.



\begin{figure}[t!]
  \centering
  \includegraphics[width=\linewidth]{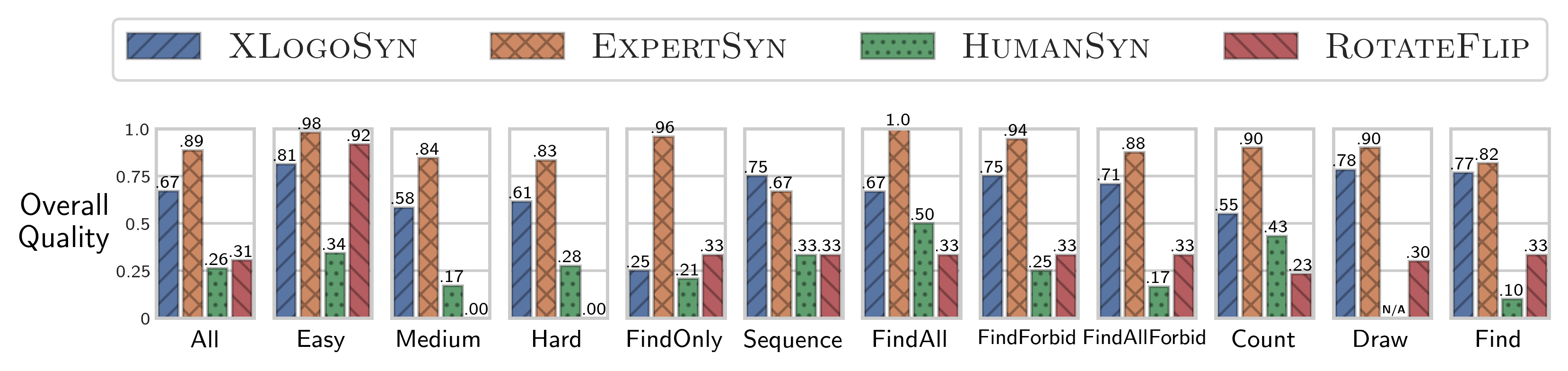} \label{fig:comparison:overall}
  \vspace{-0.6cm}
  \caption{The performance of our technique \ours{} and three baseline techniques. On the x-axis, we present the aggregated results over all scenarios (All), followed by aggregated results based on task difficulty (\Easy, \Medium, and \Hard) and based on $8$ task types. The y-axis presents the score for \emph{overall quality}. \ours{} demonstrates performance close to \expertsyn{}, and surpasses both \humansyn{} and \rotateflip{}.
  }
\label{fig:comparison}
\end{figure}


{\textbf{Results.}} We first checked the inter-rater reliability of the two human evaluators using the quadratic-weighted Cohen's kappa score~\cite{Cohen1960ACO}, achieving a near-perfect agreement of $0.84$. Next, we compare \ours{}'s performance w.r.t. baseline techniques and report statistical significance using $\chi^2$-test~\cite{cochran1952chi2}. The results are shown in Fig.~\ref{fig:comparison}. \ours{} has an overall quality score of 0.67 and is: 
(i) significantly lower than \expertsyn{} that has a score of $0.89$ ($\chi^2 = 38.8; p < 0.01$); 
(ii) significantly higher than \humansyn{} that has a score of $0.26$ ($\chi^2 = 75.5; p < 0.01$); and (iii) significantly higher than \rotateflip{} that has a score of $0.31$ ($\chi^2 = 125.2; p < 0.01$). 
\looseness-1All techniques show a performance decline with increasing task difficulty, indicating that generating more difficult tasks remains challenging for all, including experts. We found that \humansyn{} struggles because users create tasks without any references and tend to incorporate diverse concepts in a task. \rotateflip{} is effective in generating \Easy{} tasks; however, its performance drops to zero for \Medium{} and \Hard{} tasks, indicating that rotations and flips are not sufficient for generating tasks of higher difficulty levels.


\section{Deployment on XLogoOnline and Initial Results}\label{sec.deployment}

\looseness-1In this section, we present the current status of our deployment and report on initial results. We have deployed the synthesized practice tasks into the XLogoOnline platform. For each reference task on the platform, \ours{} synthesized $10$ tasks across three difficulty levels: $3$ \Easy{}, $4$ \Medium{}, and $3$ \Hard{}. After attempting a reference task, learners can choose to attempt the synthesized practice tasks or move to the \emph{next reference task} by clicking the ``Next'' button on the platform. 
Preliminary statistics, based on data collected on XLogoOnline from November 2023 to March 2024, show around $13,000$ visits on the platform and over $600,000$  execution attempts to solve tasks. Out of these attempts, $87\%$ were on reference tasks, and $13\%$ were on practice tasks synthesized by our technique.

Next, we analyze the data to investigate the educational benefits of synthesized practice tasks. We aim to answer the following research question: \emph{Do synthesized practice tasks enhance learners' success rates on the next reference tasks?} To this end, we analyze two groups: 
(i) The first group consisted of learners who failed a reference task and then moved directly to the next reference task without attempting any synthesized practice tasks; 
(ii) The second group consisted of learners who failed a reference task, then attempted the synthesized practice tasks of this reference task, and finally moved to the next reference task.

For comparison, we define the \emph{success rate} of a group of learners w.r.t a task as the percentage of learners who successfully solved a task at least once. We calculate success rates for both groups across $36$ pairs of consecutive reference tasks on the platform. The first group comprises $4,477$ learners, with a success rate of $49.2\%$. The second group includes $75$ learners, with a higher success rate of $68.0\%$. These initial results indicate that synthesized practice tasks have the potential to enhance success rates on the next reference task.


\section{Limitations and Future Work}\label{sec.discussion}

\looseness-1In this section, we discuss some limitations of our current work and ideas to tackle them in the future. First, we specify the difficulty of the synthesized tasks using pre-defined rules, which may not align with learners' perception of task difficulty. In the future, it would be important to derive a more refined notion of task difficulty by analyzing learners' interactions with the platform. Second, \ours{} does not incorporate a learner's code during task synthesis, which limits its effectiveness in personalizing practice tasks. It would be interesting to extend our technique to generate tasks personalized to the learner's misconceptions on the platform. Third, in our current implementation, generating a single high-quality task using our technique is time-consuming as it requires synthesizing and selecting from a large pool of tasks. In future work, it would be interesting to develop learning-based strategies and explore generative AI models to accelerate the synthesis process while maintaining the high quality of the synthesized tasks.


\paragraph{\textbf{Acknowledgments.}} Funded/Co-funded by the European Union (ERC, TOPS, 101039090). Views and opinions expressed are however those of the author(s) only and do not necessarily reflect those of the European Union or the European Research Council. Neither the European Union nor the granting authority can be held responsible for them.

\bibliographystyle{plain}
\bibliography{main}
%

\end{document}